# Chemical state analysis of reactively sputtered zinc vanadium nitride: The Auger parameter as a tool in materials design

Siarhei Zhuk, Sebastian Siol[*]

Empa – Swiss Federal Laboratories for Materials Science and Technology,
8600 Dübendorf, Switzerland

[*]E-mail: sebastian.siol@empa.ch

**Abstract**

Photoelectron spectroscopy is an important tool for the development of new materials. However, especially for nitride semiconductors, the formation of surface oxides, surface band bending as well as the lack of a suitable charge reference often prevent a robust analysis. Here, we perform a comprehensive chemical state analysis of the Zn-V-N phase space using the Auger parameter concept, which is less sensitive to such uncertainties. Phase-pure $Zn_2VN_3$, VN, and $Zn_3N_2$ samples are analyzed using XPS/HAXPES after transfer in inert-gas atmosphere. In addition, high-throughput chemical state analysis is performed on combinatorial $Zn_{1-x}V_xN$ thin film libraries. The evolution of the Zn Auger parameter in $Zn_{1-x}V_xN$ is consistent with previous mapping of the structural and functional properties. Strikingly, the study reveals a narrower stability range of wurtzite $Zn_{1-x}V_xN$ than our previous high-throughput XRD screening, highlighting the sensitivity of the measurement approach. The procedures applied here are transferable to many other material systems and could be particularly useful for the high-throughput development of materials with low crystallinity where insights from XRD screenings are limited.

## 1. Introduction

The development of novel functional semiconducting materials is critical for continuing progress in cutting-edge electronic technologies. Fueled by advances in computational prediction, as well as high-throughput experimental screening, novel materials are being discovered at an ever-increasing rate.[1–4] Combinatorial synthesis and high-throughput characterization methods play an important part in the accelerated exploration of complex phase spaces.[5–10] Nitrides in particular show a high innovation potential with several new stable and metastable phases reported every year. [3,4,11–14] X-ray photoelectron spectroscopy (XPS) is an important method for the discovery and design of new semiconducting phases revealing information about their electronic structure, surface composition and ideally chemical state of the constituting elements.[15–17] Chemical state analysis can provide important and complementary insights into the phase formation of novel materials.[18–20] Despite the potential of chemical state analysis for Materials Design, most high-throughput XPS studies to date focus on measurements of the surface composition.[21–23]

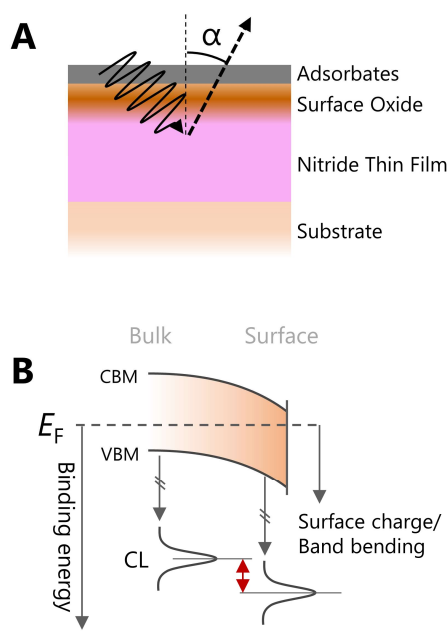

**Figure 1.** (a) Schematic structure of nitride thin film sample with surficial oxide and adsorbates used for XPS analysis. The information depth λ·cos(α) can be increased by measuring at shallow take-off angles. (b) Effect of surface band bending on the observed binding energy of core level lines.

There are several challenges associated with the chemical state analysis via XPS. Especially when analyzing nitride thin films (see **Figure 1**). The most common issues are: (I) surface contamination or surface oxidation of ex-situ treated samples, (II) the lack of a suitable charge reference, and (III) surface band bending (see **Figure 1**).[24,25] The latter is particularly important for semiconductors. Here differential charging as well as X-ray induced surface and/or junction photo voltage can alter the Fermi level position at the surface, and consequently the observed binding energies significantly (see **Figure 1B**).[26–29] Regarding the potential presence of surface oxides, it is worth mentioning that in-situ sputter cleaning of the surface using Ar$^+$ ions in the XPS chamber is typically not suitable for nitrides due to an immediate modification of the surface chemistry.[30]

Each of these challenges can be addressed, one way or another. On one hand, the effect of surface contamination can be reduced by increasing the probing depth, including usage of hard X-ray photoelectron spectroscopy (HAXPES), analysis at shallow take-off angles, or analysis of shallow core levels with lower binding energies. [17,29,31,32] On the other hand, studies of the modified Auger parameter (AP) provide a way to differentiate surficial oxide and the underlying phases as Auger lines are very sensitive to changes in local chemical environment.[19,33,34] The AP is defined as α′ = $CL_{BE}$ + $AE_{KE}$, where $CL_{BE}$ is the binding energy of the core level photoelectron and $AE_{KE}$ is the kinetic energy of Auger electron.[19] The AP is insensitive to charging effects, which can alleviate the problem of surface band bending, when the core level and Auger electrons originate from a similar probing depth.[19,29] In addition, the AP is not affected by improper charge correction, which helps to avoid uncertainties in the position of the Fermi level and allows comparing XPS results from different instruments and under different measurement conditions.[29] This ability to minimize the measurement uncertainty makes the AP a great tool for the study of insulating or semiconducting materials.

Recently we reported on the discovery and synthesis of $Zn_2VN_3$ a wide bandgap p-type semiconductor in disordered wurtzite structure.[35] We found that $Zn_2VN_3$ is an interesting material for optoelectronic applications, exhibiting room temperature photoluminescence at 2.5 eV, p-type conductivity and relatively high Hall mobility of 80 cm$^2$/V·s despite exhibiting cation disorder. As part of this work, we performed a comprehensive screening of the Zn-V-N phase space using combinatorial reactive co-sputtering combined with automated analysis. In the present manuscript, we will use a combination of the measurement approaches outlined above to perform a robust chemical state analysis on these zinc vanadium nitride thin films. First, we perform a detailed chemical state analysis using a combination of XPS and HAXPES techniques on reactively sputtered single-phase $Zn_2VN_3$, $Zn_3N_2$ and VN thin films via inert gas transfers from the synthesis to the measurement equipment. Subsequently, we use the insights from this analysis to analyze ex-situ treated combinatorial Zn-V-N libraries to investigate the evolution of the chemical state (i.e. the Auger parameter) as a function of the cation concentration and synthesis parameters. We find that the results from the chemical state analysis correlate strongly with results from X-ray diffraction mapping as well as the evolution of the functional properties in this material system, highlighting the merit of combinatorial chemical state analysis using the Auger parameter concept for Materials Discovery and Design. Finally, the approaches and concepts introduced in this paper are applicable to many other material systems where surface oxidation and surface band bending play a role.

## 2. Experimental methods

AJA 1500F sputtering system was used to synthesize single-phase $Zn_2VN_3$, $Zn_3N_2$ and VN thin films as well as combinatorial sample libraries of $Zn_{1-x}V_xN$. Phase-pure $Zn_3N_2$ and VN samples were deposited onto pre-cleaned 1.1 mm thick borosilicate glass substrates while $Zn_2VN_3$ thin films were grown epitaxially on an α-$Al_2O_3$ (0001) substrate. Radio-frequency (RF) co-sputtering method from Zn and V targets (2'' diameter and >99.9% purity) was carried out in reactive atmosphere of argon (12 sccm) and nitrogen (18 sccm). A deposition pressure of 0.5 Pa and 1 Pa was used for synthesis of single-phase and combinatorial samples, respectively. Nitrogen gas was supplied directly to the chimneys of the guns to facilitate dissociation and enhance the chemical reactivity of the process gas. For the fabrication of combinatorial $Zn_{1-x}V_xN$ samples, the deposition was performed on borosilicate glass substrates (50.8 mm x 50.8 mm) without substrate rotation to produce sample libraries with varying alloying concentration as reported elsewhere.[35] An orthogonal temperature gradient was applied using a custom sample holder design, resulting in substrate temperatures in the range from 114°C to 220°C. A Bruker D8 X-ray diffraction (XRD) system was used for automated θ-2θ mapping of structural properties of the synthesized thin films. The measurement was carried out in Bragg-Brentano geometry using a Cu kα radiation source with Ni filter. Composition at the sample libraries was characterized using a Bruker M4 Tornado X-ray fluorescence (XRF) system with Rh X-ray radiation. For the surface analysis on single-phase $Zn_2VN_3$, $Zn_3N_2$ and VN, the samples were transferred in $N_2$ atmosphere from the sputtering chamber to a glovebox connected to the XPS/HAXPES system. Prior to the analysis, the samples were stored in this glove box (Ar with < 5 ppm $H_2O/O_2$) for several days and then transferred in UHV into the HAXPES system. A PHI Quantes photoelectron spectroscopy system equipped with both hard Cr kα and Al kα X-ray radiation sources was used for the analysis of the single-phase samples. The measurement was performed at photoelectron take-off angle of 5° (see **Table S1**). A PHI Quantera system equipped with an Al kα monochromatic X-ray radiation source was employed for automated measurements of combinatorial sample libraries. The measurements were performed at a pressure below 2·$10^{-6}$ Pa. The characterization was carried out at an electron take-off angle of 45°, while power of 50W and voltage of 15 KeV were used for electron beam generation. The linearity of the energy scale was calibrated using the Au $4f_{7/2}$ (83.96 eV) and Cu $2p_{3/2}$ (932.62 eV) lines for the XPS instrument and Au $4f_{7/2}$ excited with both Cr kα and Al kα for the HAXPES instrument. Charge neutralization was achieved using a low-energy electron flood gun. C 1s at 284.8 eV was used as a charge reference, resulting in a typical measurement error of ±0.2 eV, which does not affect the values for the modified Auger parameter. Peak fitting was performed after Shirley background subtraction using Voigt profile with GL ratios of ~30 and ~70 for the XPS and HAXPES measurements, respectively. The Auger emission lines were assigned based on the work of Coghlan and Clausing.[36] The inelastic mean free path λ (IMFP) of photoelectrons in $Zn_2VN_3$ was calculated using the Quases software package, based on the Tanuma, Powell, Penn formula (TPP2M).[37] At a take-off angle α the probing depth can be calculated via 3 × λ cos(α).

## 3. Results and discussion

### 3.1 Chemical state analysis of zinc vanadium nitride on single-phase samples

In the first part of this study, we perform detailed surface analysis on selected reference samples, including single-phase $Zn_2VN_3$, $Zn_3N_2$ as well as VN. **Figure 2** shows XRD patterns of $Zn_2VN_3$, $Zn_3N_2$, and VN thin films along with their crystal structures. The obtained data are consistent with XRD reference patterns from literature.[35,38] Although the XRD patterns of $Zn_2VN_3$ and $Zn_3N_2$ are similar, $Zn_2VN_3$ demonstrates a characteristic peak at 47.6°, which is attributed to the (102) plane. Later in the study, the intensity of this peak will be used as a proxy for the $Zn_2VN_3$ content in the combinatorial $Zn_{1-x}V_xN$ thin film libraries. In our earlier study, we reported that our single-phase $Zn_2VN_3$ films exhibit a stoichiometric composition with a total amount of oxygen and carbon contamination significantly less than 1 at.%.[35] However, samples exposed to ambient conditions were found to exhibit a surface oxide layer. For this part of the study, the samples were transferred in inert-gas atmosphere into the XPS/HAXPES system to minimized surface oxidation.

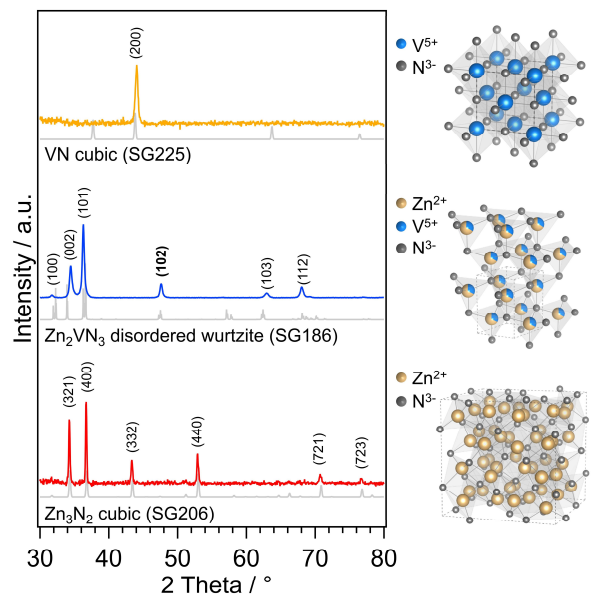

**Figure 2.** XRD patterns of $Zn_3N_2$, $Zn_2VN_3$ and VN thin films synthesized via reactive co-sputtering on borosilicate glass substrates. The (102)-reflection of $Zn_2VN_3$ is unique among the three phases. The unit cell structures are given for the respective phases.

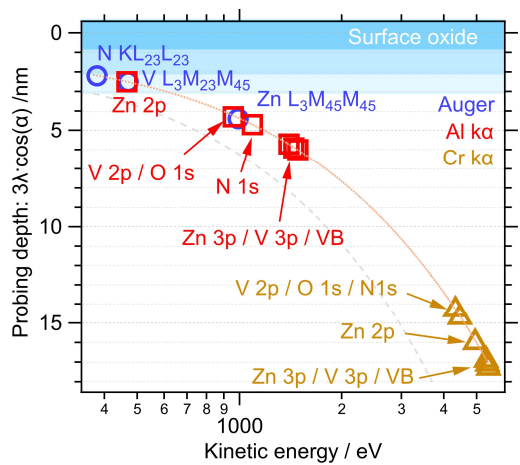

**Figure 3.** Probing depth (3 × λcos(α)) of the relevant photoemission and Auger lines in $Zn_2VN_3$ calculated for Al kα and Cr kα X-ray excitation sources (photoelectron take-off angle of α = 45°). The gray dashed line represents calculated information depth for measurements conducted at photoelectron take-off angle of α = 5°.

When performing chemical state analysis, especially when soft and hard X-rays are being used, the different probing depths associated with different core level- and Auger emission lines need to be considered.[29] **Figure 3** and **Table S1** summarizes information depths of photoemission and Auger lines for elements constituting $Zn_2VN_3$ as a function of their kinetic energy of photoelectrons and Auger electrons. In addition, the figure illustrates the increased probing depth when low take-off angles are being used. A detailed chemical state analysis of $Zn_2VN_3$ includes Zn $L_3M_{45}M_{45}$, V $L_3M_{23}M_{45}$ and N $KL_{23}L_{23}$ Auger electron emission spectra as well as the corresponding core levels. The information depth of the Auger emission is independent of the excitation source and consequently cannot be increased using hard X-rays. In order to maximize the probing depth a low photoelectron take-off angle of 5° was used for the XPS/HAXPES analysis on the single-phase samples. In addition, it is important to match the Auger emission with a core level emission of similar probing depth to avoid errors due to differential charging. In practice, this means using core-levels measured with soft X-rays, rather than hard X-rays. However, HAXPES measurements in our case can provide insights into whether observed oxide phases are localized at the surface or not.

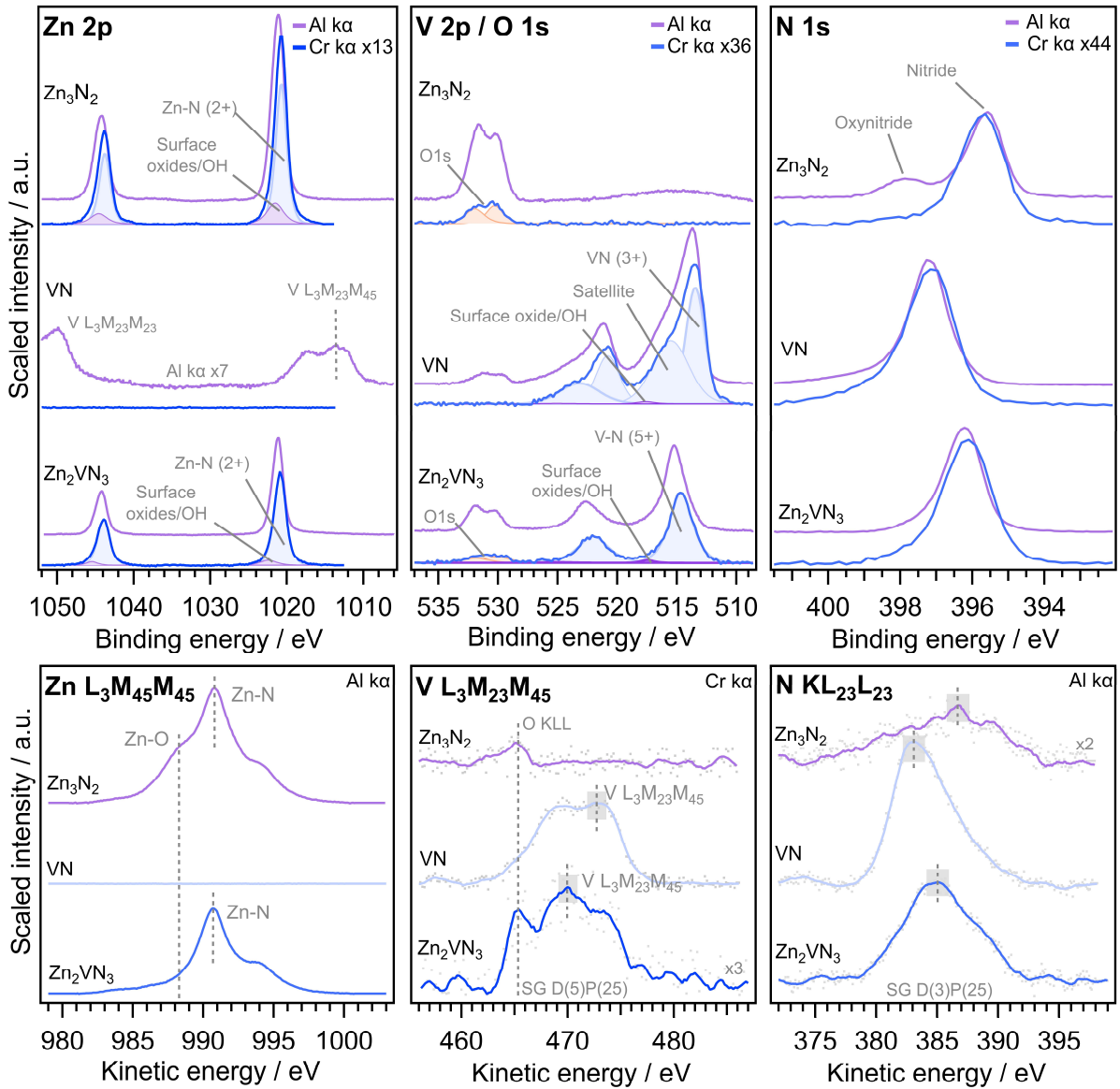

**Figure 4.** XPS/HAXPES analysis of Zn 2p, V 2p, N1s core levels as well as Zn $L_3M_{45}M_{45}$, V $L_3M_{23}M_{45}$, and N $KL_{23}L_{23}$ Auger electron emission lines of single-phase $Zn_2VN_3$, $Zn_3N_2$ and VN thin films transferred in inert gas conditions. V $L_3M_{23}M_{45}$ and N $KL_{23}L_{23}$ were smoothed (SG D(5)P(25) and SG D(3)P(25), respectively) due to their low signal to noise ratios. For $Zn_2VN_3$ the V Auger emission overlaps with the Zn 2p core level region when measured with Al kα, hence the V $L_3M_{23}M_{45}$ was excited using Cr kα radiation.

**Figure 4** shows a full set of spectra for epitaxially stabilized $Zn_2VN_3$ as well as $Zn_3N_2$ and VN thin films using combination of XPS and HAXPES. Spectra measured using Cr kα radiation source are scaled due to the difference in photoelectron ionization cross-section relative that of Al kα. Despite using an inert gas transfer some surface oxidation is visible in the core levels measured with Al kα. As shown in **Figure 3**, the Zn 2p core level is very surface sensitive. Therefore, it is critical to perform peaks fitting for the measured spectra. Based on the intensity of the surface oxide component in the Zn 2p as well as the O 1s intensity, it can be seen that $Zn_2VN_3$ is less prone to oxidation relative to $Zn_3N_2$. For the Cr kα the intensity of the surface oxide component is much lower, indicating that the oxide is indeed localized at the samples surface. This observation is also consistent with the data obtained for Zn $L_3M_{45}M_{45}$. The information depth for Zn $L_3M_{45}M_{45}$ is higher than for Zn 2p (see **Figure 3**). In addition, unlike in the Zn 2p core level emission, the Zn-O component is clearly distinct from the Zn-N component in the Zn $L_3M_{45}M_{45}$ spectra. This makes the Zn $L_3M_{45}M_{45}$ Auger line ideally suited for Auger parameter studies. Similar to Zn 2p, the spectra for V 2p measured with the Cr kα source show almost no contribution from surface oxides. Noteworthy is the asymmetric peak shape in the case of VN which has been attributed to the presence of shake-up satellite peaks. These satellite peaks are reported to originate from poorly screened core hole states.[39,40] Unfortunately the V $L_3M_{23}M_{45}$ Auger spectrum overlaps with that of Zn 2p when measured with Al kα.[41] Therefore, the Auger line was excited using Cr kα radiation, instead. This shifts the kinetic energy of the Zn 2p core level emission and consequently isolates the V $L_3M_{23}M_{45}$ Auger line. In all other cases, exciting the Auger emission with Al kα is preferred as (i) information depth of Auger lines does not depend on X-ray excitation source and (ii) intensity of the measured signal is higher for Al kα due to the higher photoionization cross sections of the Al kα radiation. In contrast to the Zn $L_3M_{45}M_{45}$ spectrum, V $L_3M_{23}M_{45}$ Auger electron emission of $Zn_2VN_3$ is broader. In addition it shows some overlap with an O KLL peak originating from surface oxides.[42] This is confirmed by measurements on $Zn_3N_2$, which does not contain any V. Similarly to V $L_3M_{23}M_{45}$, the N $KL_{23}L_{23}$ line is also very surface sensitive with a probing depth of only about 3 nm and is therefore prone to uncertainty due to surface oxide formation. Here, the lowest signal intensity was recorded for $Zn_3N_2$, which tends to oxidize more relative to the other compounds. In fact, $Zn_3N_2$ samples stored in ambient conditions did not exhibit any N $KL_{23}L_{23}$ signal at all (not shown here). While the measurements of the N $KL_{23}L_{23}$ were successful for the samples handled in inert gas, these results also demonstrate that ex-situ analysis of the N chemical state in Zn-V-N is not feasible.

Another noteworthy observation is a slight shift to lower binding energy for most of the samples when measured using HAXPES. Shifts to higher binding energies for more surface sensitive measurements (i.e. Al kα vs. Cr kα) are a common observation for films exhibiting differential charging. The obtained XPS experimental data on chemical state analysis of $Zn_3N_2$, $Zn_2VN_3$ and VN thin films is summarized in **Table 1**. As differential charging was observed, the AP is calculated only using core level binding energies from Al kα measurements combined with the respective Auger electron kinetic energies. The full data set obtained using XPS and HAXPES analysis is presented in **Table S1**. The Zn$_3$N$_2$ AP was previously reported by Futsuhara *et al.* for Zn$_3$N$_2$ samples exposed to ambient conditions. Here the Zn 2p/Zn L$_3$M$_{45}$M$_{45}$ AP is reported slightly higher as 2012.3 eV, which is consistent with our results from ex-situ treated samples.[43] For VN no previous reports of the AP were found.

**Table 1**: XPS data on core level binding energies CL$_{BE}$ as well as Auger electron kinetic energies AE$_{KE}$ for phase-pure Zn$_3$N$_2$, Zn$_2$VN$_3$ and VN thin films transferred in inert gas conditions. Auger parameter is calculated as AP= CL$_{BE}$ + AE$_{KE}$. Al kα X-ray radiation was used for all measurements, with the exception of the V L$_3$M$_{23}$M$_{45}$ Auger emission in Zn$_2$VN$_3$, which was excited using Cr kα. The typical accuracy in determining the CL$_{BE}$ and AE$_{KE}$ better than ± 0.1 eV. For V L$_3$M$_{23}$M$_{45}$ and N KL$_{23}$L$_{23}$ the error is larger with an uncertainty in the range of ± 1.0 eV. C1s charge correction is used resulting in a typical additional uncertainty of up to ±0.2 eV. This uncertainty does not affect the AP.

| | Zn 2p$_{3/2}$ CL$_{BE}$ /eV | Zn (2p$_{3/2}$–2p$_{1/2}$) /eV | Zn 3p$_{3/2}$ CL$_{BE}$ /eV | Zn (3p$_{3/2}$–3p$_{1/2}$) /eV | V 2p$_{3/2}$ CL$_{BE}$ /eV | V (2p$_{3/2}$–2p$_{1/2}$) /eV | N 1s CL$_{BE}$ /eV | Zn L$_3$M$_{45}$M$_{45}$ AE$_{KE}$ /eV | V L$_3$M$_{23}$M$_{45}$ AE$_{KE}$ /eV | N KL$_{23}$L$_{23}$ AE$_{KE}$ /eV | Zn AP (Zn 2p) /eV | Zn AP (Zn 3p) /eV | V AP /eV | N AP /eV |
|---|---|---|---|---|---|---|---|---|---|---|---|---|---|---|
| **Zn$_2$VN$_3$** | 1021.05 | 23.05 | 88.09 | 3.00 | 514.97 | 7.35 | 396.21 | 990.72 | 469.98 | 385.06 | 2011.77 | 1078.81 | 984.95* | 781.27 |
| **Zn$_3$N$_2$** | 1020.82 | 23.05 | 87.89 | 3.00 | - | - | 395.63 | 990.81 | - | 386.73 | 2011.63 | 1078.70 | - | 782.36 |
| **VN** | - | - | - | - | 513.69 | 7.35 | 397.23 | - | 473.13 | 383.11 | - | - | 986.82 | 780.34 |

The results from the study are summarized in the form of Wagner plots (see **Figure 5**). Wagner plots can be used to illustrate experimental information on photoemission and Auger lines of different elements along with calculated AP.[34,44] **Figure 5a** shows Wagner plot for Zn using the Zn 2p core level and Zn L$_3$M$_{45}$M$_{45}$ Auger line. Although Zn 2p core level spectra of ZnO and Zn$_3$N$_2$ phases are very close in binding energy, on the Wagner plot metallic Zn, Zn$_3$N$_2$ and ZnO are easily distinguished. In the current study, presenting the data in the form of Wagner plots is particularly useful for Zn 2p as it helps to clearly differentiate the nitride and oxide phases. Note that the spread in binding energy is highlighted in green while that of AP is shown in red. As it is evident from the Wagner plot Zn$_3$N$_2$ and Zn$_2$VN$_3$ phases exhibit similar AP values. This is expected as both compounds exhibit a 4-fold coordination where Zn is present in 2$^+$ oxidation state and each Zn atom in the unit cell is surrounded by four nitrogen atoms in corner sharing tetrahedrals (Crystal structures of Zn$_3$N$_2$ and Zn$_2$VN$_3$ are given in **Figure 2**). **Figure 5b** shows the Wagner plot for V. In contrast to Zn 2p the V 2p core level demonstrates a large spread in binding energies for V, V$_2$O$_5$, VN and Zn$_2$VN$_3$ phases.

However, the calculated spacing in AP for these compounds is relatively small. Consequently, a study of the V 2p core level binding energy might be preferred over the AP. This is because the changes in binding energy are large enough to exceed typical inaccuracies in the measurements outlined above. From the V 2p binding energy alone, it is possible to distinguish the different nitride and oxide phases. It is noteworthy, that the binding energy of V $5^+$ in $Zn_2VN_3$ the nitride is dissimilar from that of $V_2O_5$. The Wagner plot for N 1s core level and N $KL_{23}L_{23}$ Auger electron emission is presented in **Figure S1**. Here, the high surface sensitivity of N $KL_{23}L_{23}$ Auger line results in noisy spectra exhibiting similar values for $Zn_2VN_3$, $Zn_3N_2$ and VN phases.

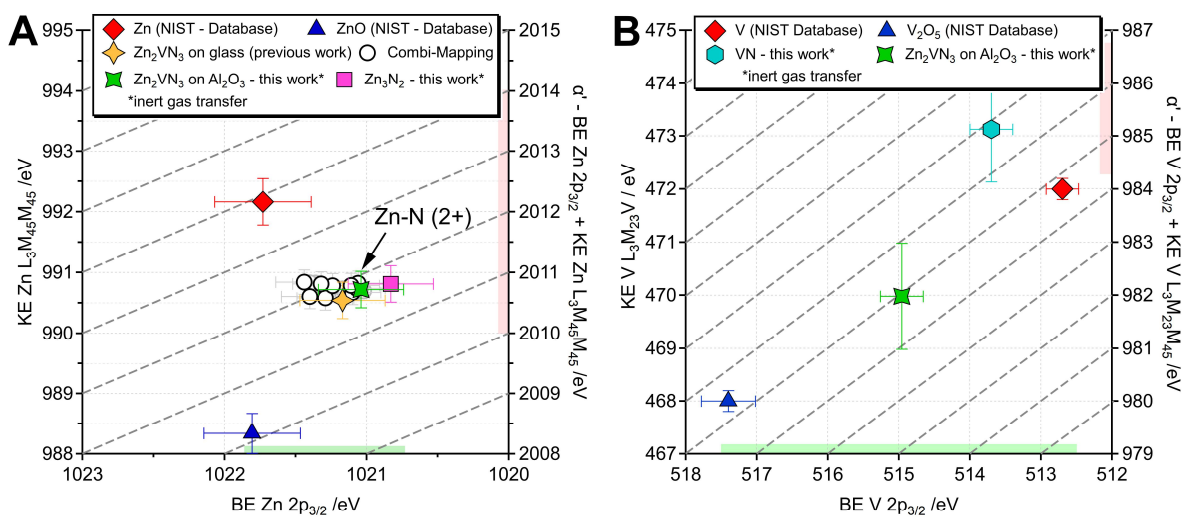

**Figure 5.** Wagner plots for (a) AP of Zn 2p core level and Zn $L_3M_{45}M_{45}$ Auger electron emission, (b) AP of V 2p core level and V $L_3M_{23}M_{45}$ Auger electron emission showing data of selected elementary, binary and ternary compounds. Zn, ZnO, V, $V_2O_5$ data are taken from NIST XPS database.[45] The AP for Zn shows a large spread for different chemical environments and allows for clear distinction of different materials. For V changes in core level binding energy are more pronounced than the corresponding spread in AP. Error bars include the uncertainty due to errors in charge correction.

### 3.2 Chemical state analysis of $Zn_{1-x}V_xN$ using XPS mapping on combinatorial libraries

Following the analysis of the single-phase samples using XPS/HAXPES, we performed chemical state analysis on combinatorial libraries using automated XPS mapping with the goal to track changes in phase formation and chemical binding environment for various alloying concentrations and deposition temperatures in $Zn_{1-x}V_xN$ thin films. Since all combinatorial sample libraries were exposed to air a different measurement strategy was used. For Zn, the AP was measured, whereas for V the V 2p core level emission was analyzed. Both V $L_3M_{23}M_{45}$ and N $KL_{23}L_{23}$ Auger electron emission lines

Chemical state analysis of reactively sputtered zinc vanadium nitride, S. Zhuk and S. Siol, 2022    11

are very surface sensitive (see **Figure 3**) and are easily affected by surface oxides surface band bending. However, the V 2p core level binding energy shows pronounced shifts, which exceed the typical errors associated with surface band bending or errors in charge correction (**Figure 5B**).

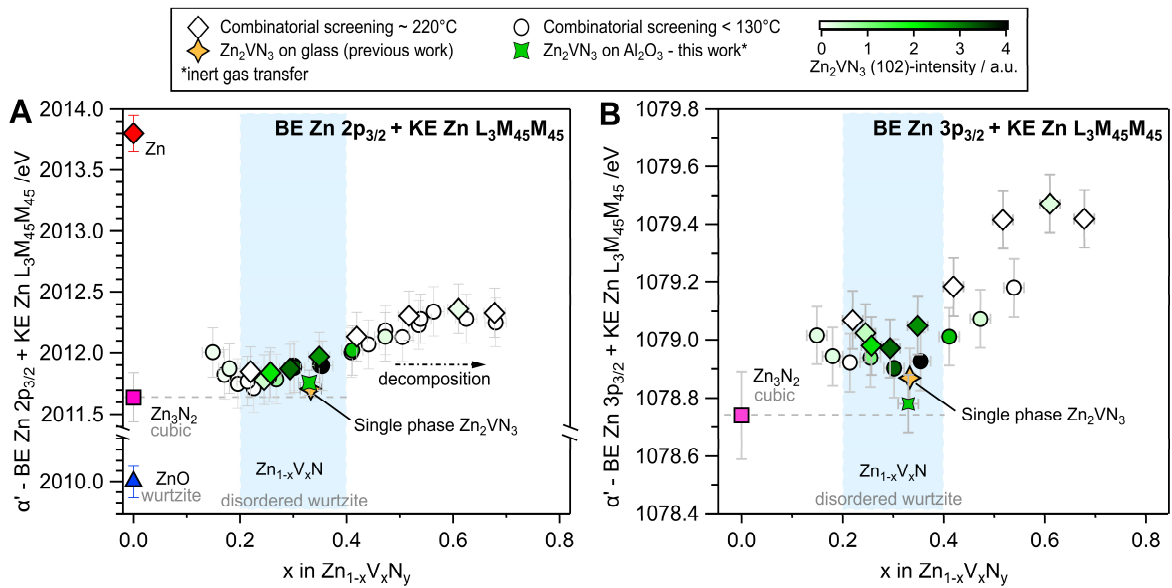

**Figure 6.** Zn AP for (a) Zn 2p and (b) Zn 3p core levels and Zn $L_3M_{45}M_{45}$ Auger electron emission plotted as a function of alloying concentration in $Zn_{1-x}V_xN$.

**Figure 6a** shows a plot of AP for Zn 2p and Zn $L_3M_{45}M_{45}$ Auger electron emission as a function of alloying concentration for combinatorial samples grown at different temperatures (~220°C and below 130°C). It can be seen that APs of phase-pure $Zn_3N_2$ and $Zn_2VN_3$ demonstrate similar values. This is explained by the nature of chemical binding in these compounds, which involves a 4-fold coordination of the Zn atom as discussed earlier. AP undergoes a pronounced shift in $Zn_{1-x}V_xN$ thin films in the range from x = 0 to x = 0.2 where content of V in the films is increased. In our recent work we determined the solubility limits for wurtzite $Zn_{1-x}V_xN$ by XRD to be between 0.2 < x < 0.4 for deposition carried out below 130 °C (blue region in **Figure 6**). Furthermore, the observed change in AP correlates well with the formation of wurtzite $Zn_{1-x}V_xN$. The correlation is highlighted by the color of the data points, which corresponds to the intensity of an XRD peak, which is only observed for the wurtzite phase (see **Figure 3**). Furthermore, at x > 0.5 decomposition of wurtzite $Zn_{1-x}V_xN$ occurs,[35] which is reflected in the increase of Zn AP value. Thus, the Zn AP for the Zn 2p core level correlates with solubility limits in $Zn_{1-x}V_xN$ while no discernible effect of synthesis temperature on evolution of Zn AP is observed. However, the relatively low kinetic energy of the Zn 2p photoelectrons makes this line very surface sensitive ($3\lambda\cos(\alpha) \approx 2.5$ nm). Therefore, surface oxidation may affect the obtained results, as the accuracy of Zn 2p components fitting is limited. For this reason, it is important to analyze a combination of the shallower Zn 3p core level (3λcos(α) ≈ 5.7 nm) with the Zn $L_3M_{45}M_{45}$ Auger line (3λcos(α) ≈ 4.4 nm). This combination provides higher and similar probing depth for both lines and consequently is much less sensitive to surface oxidation and differential charging (see **Figure 3**). Strikingly, this is reflected in a higher sensitivity to changes in the chemical state. **Figure 6b** shows the correlation between $Zn_2VN_3$ phase purity and the Zn 3p AP. A local minimum of the Zn 3p AP is observed for phase-pure $Zn_2VN_3$ thin films grown on glass and sapphire substrates which is in good agreement with reference values for single-phase $Zn_3N_2$. Additionally, the effect of deposition temperature on evolution of Zn AP with varying alloying concentration can be clearly seen now in contrast to the case of Zn 2p study. It has been found that solubility in $Zn_{1-x}V_xN$ thin films is relatively low at high synthesis temperature. For x > 0.5 increase of Zn AP is associated with decomposition of the wurtzite phase. The deviation in Zn AP from reference value is in line with the formation of VN precipitates as reported in our earlier study. Overall, the Zn 3p AP appears to be more sensitive to decomposition than our recent XRD mapping experiments.

These findings align well with our analysis of the V chemical state. **Figure 7a** shows the evolution of the V 2p binding energy as a function of alloying concentration in $Zn_{1-x}V_xN$. In order to avoid systematic measurement errors we intentionally chose the spectral regions intensity maximum rather than using a binding energy derived from a multi-component peak fitting. The study confirms that the oxidation state of V is $5^+$, as expected for $Zn_2VN_3$. It can be seen that for compositions x < 0.5 in $Zn_{1-x}V_xN$ the binding energy of V 2p is very similar to that of single phase $Zn_2VN_3$. Strikingly, the binding energy does not appear to change when increasing the alloying concentration. This indicates that V does not get incorporated in $4^+$ oxidation state which would be necessary to form the metastable $ZnVN_2$. Incorporating V in crystal lattice with oxidation state of +4 would violate the octet rule, which is also reflected in the high formation enthalpy of $ZnVN_2$. This is in line with our previous observations, that $ZnVN_2$ was not synthesized on the combinatorial libraries.[35] As the composition exceeds x = 0.5, precipitation of VN can be observed in XRD,[35] which is also reflected in a shift to lower binding energies, as expected. The VN content was further fitted in $Zn_{1-x}V_xN$ using the envelope of single-phase VN spectrum (see **Figure S3**). **Figure 7b** nicely illustrates the formation of VN precipicates once the solubility limit is exceeded. The lower non-equilibrium solubility in $Zn_{1-x}V_xN$ at higher synthesis temperatures is also clearly visible and consistent with observations made in the Zn AP study (see **Figure 6**).

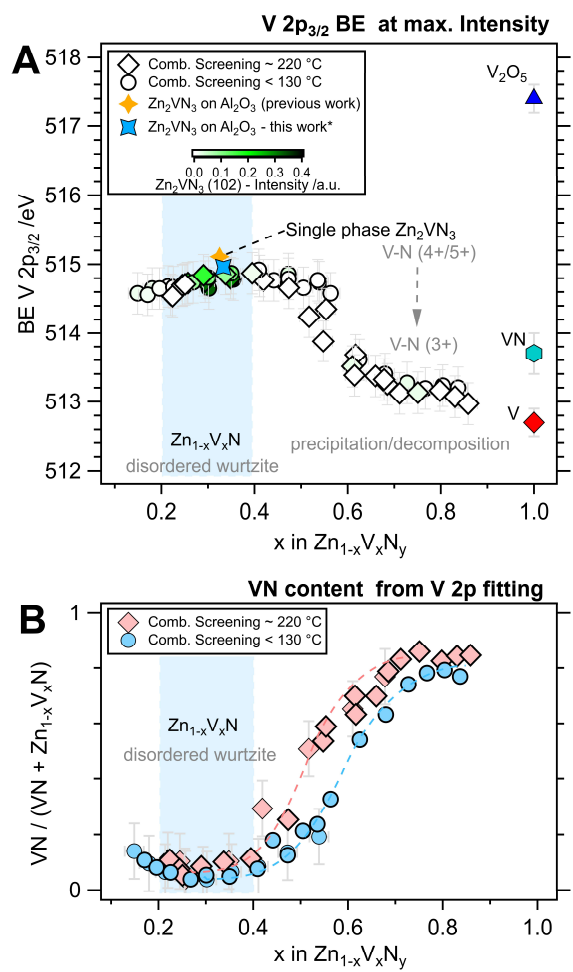

**Figure 7.** (a) Binding energy of V 2p photoelectrons and (b) VN/(VN + $Zn_{1-x}V_xN$) ratio as a function of alloying concentration in $Zn_{1-x}V_xN$.

## 4. Summary & conclusions

We demonstrate a comprehensive chemical state analysis on reactively sputtered thin films in the Zn-V-N material system. A detailed analysis is performed on the novel wide band gap semiconductor $Zn_2VN_3$ as well as VN and $Zn_3N_2$ using XPS/HAXPES and inert gas transfers. In addition, we perform high-throughput Auger parameter analysis on combinatorial libraries of $Zn_{1-x}V_xN$ to investigate the evolution of the chemical state as function of the synthesis parameters. The results are in line with our previous characterization results of the material and provide additional important insights. Specifically for phase-pure $Zn_2VN_3$, Zn and V are present in a $2^+$ and $5^+$ oxidation states, respectively. In addition, the evolution of the Zn AP as well as multi-component peak fitting of the V 2p core level indicate slightly narrower solubility limits for wurtzite $Zn_{1-x}V_xN$ than previously determined by XRD mapping. These results highlight that chemical state analysis based on the Auger parameter concept is an important complementary tool for development of semiconductor materials. If good reference data sets are available and if the probing depth of photoelectrons is considered, Auger parameter studies can even be used for analysis of ex-situ combinatorial libraries and can provide a powerful, complementary tool for materials discover and design. The measurement concepts highlighted in this study are applicable to many other materials systems. For instance, the combination of Zn 3p core level and Zn $L_3M_{45}M_{45}$ Auger electron emission is found to be particularly useful for high-throughput chemical state analysis of air exposed zinc nitrides.

**Acknowledgements**

S.Z. gratefully acknowledges financial support from the EMPA internal research call 2020. The authors acknowledge Swiss National Science Foundation (R'Equip program, Proposal No. 206021_182987) for providing funds for the acquisition of the PHI Quantes XPS/HAXPES system. In addition, the authors would like to thank Claudia Cancellieri for support with the sample transfer and during the HAXPES experiments.

Supplementary information for

# Chemical state analysis of reactively sputtered zinc vanadium nitride: The Auger parameter as a tool in materials design

Siarhei Zhuk, Sebastian Siol[*]

Empa – Swiss Federal Laboratories for Materials Science and Technology,
8600 Dübendorf, Switzerland

[*]E-mail: sebastian.siol@empa.ch

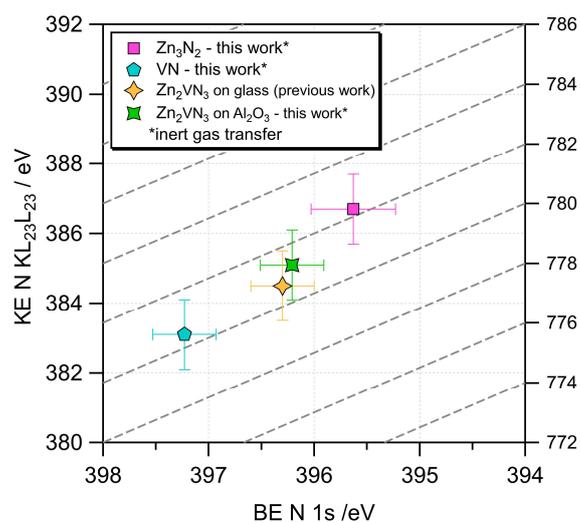

**Figure S1.** Wagner plots for (a) AP of N1s core level and N $KL_{23}L_{23}$ Auger electron emission for selected elementary, binary and ternary nitride compounds.

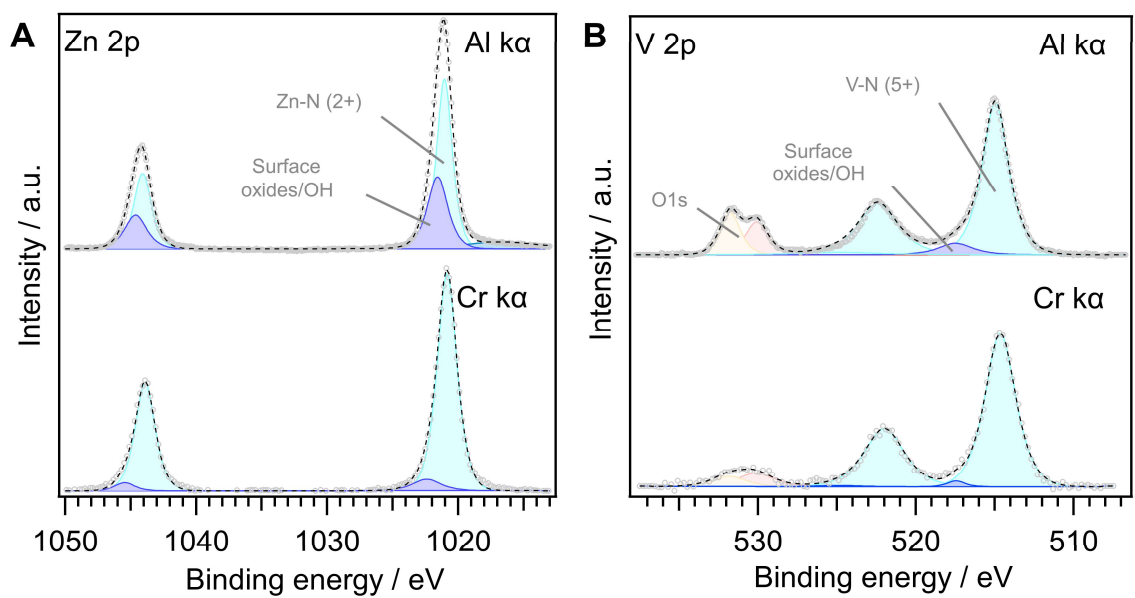

**Figure S2.** XPS/HAXPES analysis of phase-pure epitaxially stabilized $Zn_2VN_3$ thin films transferred in inert gas conditions.

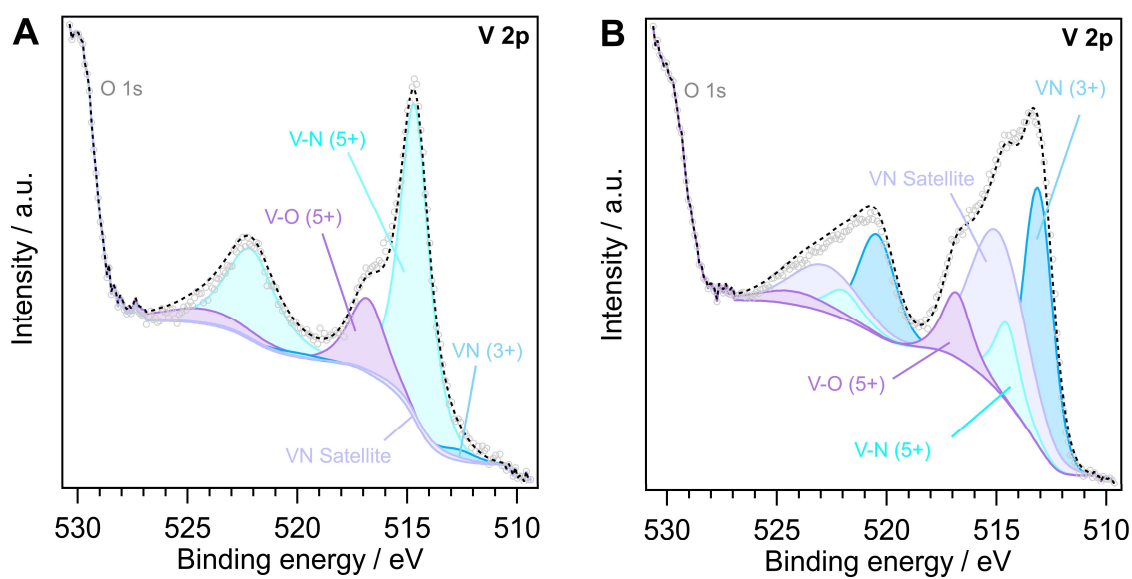

**Figure S3.** Example of the peak fitting performed on V 2p core level spectra for ex-situ combinatorial sample $Zn_{1-x}V_xN$ with composition for two representative compositions (a) x ~0.33 and (b) x ~0.5 deposited at $T < 130°C$.

**Table S1:** Probing depth for various regions related to Zn-V-N material system measured using XPS and HAXPES at an electron take-off angle of 5° and 45° to the substrate normal.

| Region | BE /eV | KE /eV | | IMFP/Å | | 3*IMFP*COS(5°) /nm | | 3*IMFP*COS(45°) /nm | |
|---|---|---|---|---|---|---|---|---|---|
| | | Al kα | Cr kα | Al kα | Cr kα | Al kα | Cr kα | Al kα | Cr kα |
| **Zn 2p** | 1021 | 465.7 | 4952 | 12.0 | 75.5 | 3.6 | 22.6 | 2.5 | 16 |
| **Zn 3p** | 88 | 1398.7 | 5329.7 | 27 | 79.9 | 8.1 | 23.9 | 5.7 | 16.9 |
| **V 2p/O 1s** | 528 | 958.7 | 4459 | 20.3 | 69.2 | 6.1 | 20.7 | 4.3 | 14.7 |
| **V 3p** | 40 | 1446.7 | 5377.7 | 27.7 | 80.5 | 8.3 | 24.1 | 5.9 | 17.1 |
| **N 1s** | 396 | 1090.7 | 4327 | 22.3 | 67.6 | 6.7 | 20.2 | 4.7 | 14.3 |
| **Zn $L_3M_{45}M_{45}$** | | 990.5 | 990.5 | 20.8 | 20.8 | 6.2 | 6.2 | 4.4 | 4.4 |
| **V $L_3M_{23}M_{45}$** | | 470 | 470 | 12 | 12 | 3.6 | 3.6 | 2.5 | 2.5 |
| **N $KL_{23}L_{23}$** | | 380 | 380 | 10.3 | 10.3 | 3.1 | 3.1 | 2.2 | 2.2 |
| **Vb** | 0 | 1486.7 | 5417.7 | 28.4 | 81.4 | 8.5 | 24.4 | 6 | 17.3 |

**Table S2**: XPS/HAXPES data on core level binding energies $CL_{BE}$ as well as Auger electron kinetic energies $AE_{KE}$ for $Zn_2VN_3$, $Zn_3N_2$ and VN thin films. The modified Auger parameter is calculated as AP= $CL_{BE}$ + $AE_{KE}$. The binding energies determined with Cr kα appear lower than for Al kα which can be explained by differential charging. Data reported as measured. The typical accuracy in determining the $CL_{BE}$ as well as $AE_{KE}$ are on the order of ± 0.1 eV, for V $L_3M_{23}M_{45}$ and N $KL_{23}L_{23}$ the measurement error is much larger with uncertainties in the range of ± 1.0 eV. C1s charge correction is used for the analysis, resulting in a typical inaccuracy of 0.2 eV. This uncertainty does not affect the AP.

| | Zn $2p_{3/2}$ $CL_{BE}$ /eV | Zn ($2p_{3/2}$–$2p_{1/2}$) /eV | Zn $3p_{3/2}$ $CL_{BE}$ /eV | Zn ($3p_{3/2}$–$3p_{1/2}$) /eV | V $2p_{3/2}$ $CL_{BE}$ /eV | V ($2p_{3/2}$–$2p_{1/2}$) /eV | N 1s $CL_{BE}$ /eV | Zn $L_3M_{45}M_{45}$ $AE_{KE}$ /eV | V $L_3M_{23}M_{45}$ $AE_{KE}$ /eV | N $KL_{23}L_{23}$ $AE_{KE}$ /eV | Zn AP (Zn 2p) /eV | Zn AP (Zn 3p) /eV | V AP /eV | N AP /eV |
|---|---|---|---|---|---|---|---|---|---|---|---|---|---|---|
| **Al kα** | | | | | | | | | | | | | | |
| **$Zn_2VN_3$** | 1021.05 | 23.05 | 88.09 | 3.00 | 514.97 | 7.35 | 396.21 | 990.72 | - | 385.06 | 2011.77 | 1078.81 | 984.95* | 781.27 |
| **$Zn_3N_2$** | 1020.82 | 23.05 | 87.89 | 3.00 | - | - | 395.63 | 990.81 | - | 386.73 | 2011.63 | 1078.70 | - | 782.36 |
| **VN** | - | - | - | - | 513.69 | 7.35 | 397.23 | - | 473.13 | 383.11 | - | - | 986.82 | 780.34 |
| **Cr kα** | | | | | | | | | | | | | | |
| **$Zn_2VN_3$** | 1020.84 | 23.05 | 87.83 | 3.00 | 514.65 | 7.35 | 396.12 | 990.74 | 469.98 | 384.67 | -+ | -+ | - | -+ |
| **$Zn_3N_2$** | 1020.72 | 23.05 | 87.77 | 3.00 | - | - | 395.68 | 990.54 | - | - | -+ | -+ | -+ | -+ |
| **VN** | - | - | | | 513.44 | 7.35 | 397.10 | - | 472.75 | 382.89 | - | - | -+ | -+ |

*Combination of Al kα core level and Cr kα Auger emission, +inaccurate due to differential charging